# Resource Allocation in Co-existing Optical Wireless HetNets

Osama Zwaid Alsulami[1], Sarah O. M. Saeed[1], Sanaa Hamid Mohamed[1],
T. E. H. El-Gorashi[1], Mohammed T. Alresheedi[2] and Jaafar M. H. Elmirghani[1]
[1]School of Electronic and Electrical Engineering, University of Leeds, LS2 9JT, United Kingdom
[2]Department of Electrical Engineering, King Saud University, Riyadh, Kingdom of Saudi Arabia
ml15ozma@leeds.ac.uk, elsoms@leeds.ac.uk, elshm@leeds.ac.uk,
t.e.h.elgorashi@leeds.ac.uk, malresheedi@ksu.edu.sa, j.m.h.elmirghani@leeds.ac.uk

**ABSTRACT**

In multi-user optical wireless communication (OWC) systems interference between users and cells can significantly affect the quality of OWC links. Thus, in this paper, a mixed-integer linear programming (MILP) model is developed to establish the optimum resource allocation in wavelength division multiple access (WDMA) optical wireless systems. Consideration is given to the optimum allocation of wavelengths and access points (APs) to each user to support multiple users in an environment where Micro, Pico and Atto Cells co-exist for downlink communication. The high directionality of light rays in small cells, such as Pico and Atto cells, can offer a very high signal to noise and interference ratio (SINR) at high data rates. Consideration is given in this work to visible light communication links which utilise four wavelengths per access point (red, green, yellow and blue) for Pico and Atto cells systems, while the Micro cell system uses an infrared (IR) transmitter. Two 10-user scenarios are considered in this work. All users in both scenarios achieve a high optical channel bandwidth beyond 7.8 GHz. In addition, all users in the two scenarios achieve high SINR beyond the threshold (15.6 dB) needed for $10^{-9}$ on off keying (OOK) bit error rate at a data rate of 7.1 Gbps.

## 1. INTRODUCTION

The number of users connected to the Internet is likely to increase significantly as a result of new technologies, such as the Internet of Things (IoT). Cisco has estimated that there will be approximately 27 times more connected users between 2016 and 2021 [1]. The capacity of the radio frequency (RF) spectrum used by the current wireless technology is insufficient to be able to handle such an increase, and thus the demand for high data rates will be a major challenge when utilising the RF spectrum. Optical wireless communication (OWC) systems have been posited as a potential solution for supporting multiple-users and high data rates, and is considered to be part of the sixth generation (6G) of communication systems. Recently, OWC systems have gained the interest of many researchers [2] – [8] because they can provide higher capacity and better security compared to RF systems [3], [4]. Moreover, OWC systems have been widely shown to support high data rates of up to 25 Gbps in indoor environments [4]–[13].

Many schemes have been proposed to improve optical wireless communication links. These include beam adaptation, which can be used in the form of a range of adaptation schemes such as angle, power and delay adaptation [3], [14]–[20]. Transmitters with multiple beams can be used to improve the signal to noise ratio by reducing the effective transmitter to receiver distance using spot diffusing approaches [21] – [25]. Furthermore, the optical wireless communication link can be improved by reducing interference by using diversity techniques, such as an angle diversity receiver (ADR) [26] - [29]. Uplink OWC systems have been proposed in different studies [30], [31], however, more consideration must be giving to energy efficiency [32]. Multiple users can be accommodated using different code division multiple access (CDMA) schemes including pulse position modulation (PPM) CDMA, and multi-carrier CDMA (MC-CDMA) [33], [34]. However, interference has a significant impact in OWC systems when multiple users operate simultaneously which affects the system's performance. Thus, different schemes based on different orthogonal resources have been considered to reduce interference by using orthogonal time, wavelength or frequency resources [13], [35]–[39]. In particular, wavelength division multiple access (WDMA) schemes can offer a potential native solution in (multi-colour) VLC systems to support multiple users in OWC systems and reduce interference between users.

This paper studies the resource allocation problem in co-existing optical wireless HetNets. Three types of communication cells have been considered in this work; namely micro, pico and atto cells [11]. Two types of transmitters are used in this work; infrared and visible light Laser Diodes (LDs) which consist of four wavelengths (red, yellow, green and blue). Therefore, WDMA is used in this paper for providing multiple access. The LDs used can provide white lighting for indoor illumination [40] and high bandwidth for communication. In addition, a 7 branch angle diversity receiver, similar to [11], is examined in this work. The optimum resource allocation of cells, access points (APs) and wavelengths is determined based on maximising all users' SINRs by developing and using a Mixed Integer Linear Programme (MILP) optimisation model similar to that in [35].

The rest of this paper is ordered as follows: the configuration of the system, including the room, cells, transmitters and receiver configuration, is presented in Section 2, while Section 3 introduces the simulation results. Finally, the conclusions are presented in Section 4.

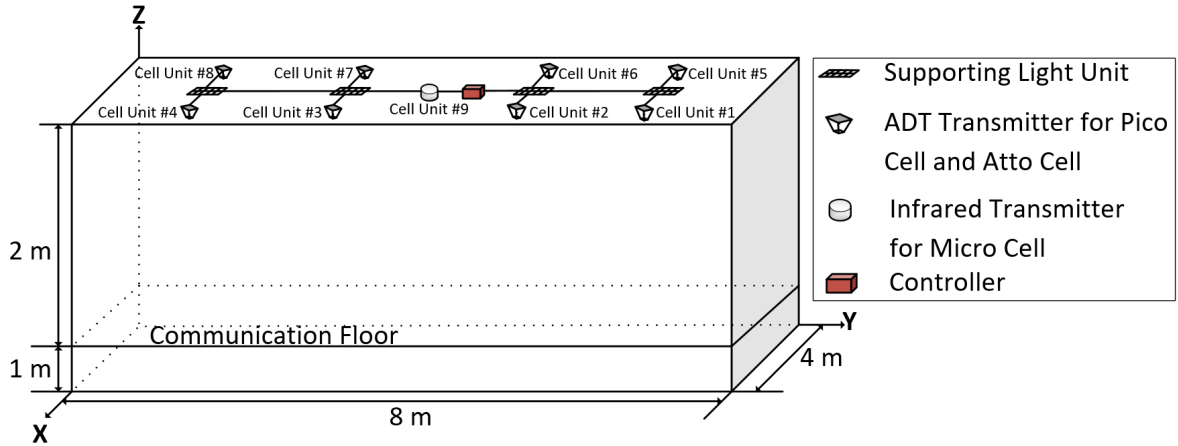

Figure 1. Room Configuration

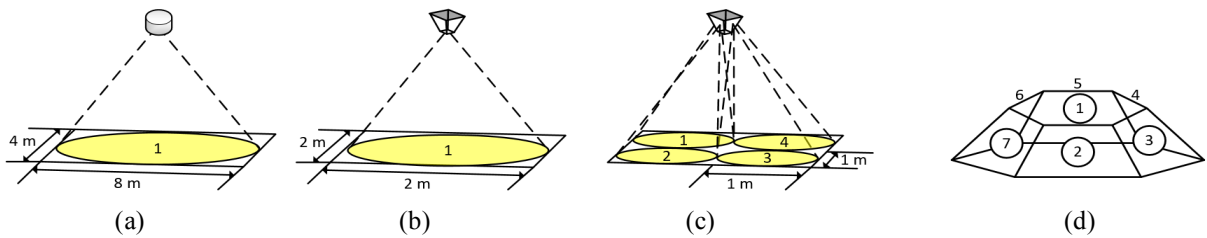

Figure 2. (a) Micro cell, (b) Pico cell, (c) Atto cell and (d) Receiver configuration.

## 2. SYSTEM CONFIGURATION

A room with no furniture that has no doors or windows has been evaluated in this work (see Figure 1). The room is supported by three cellular optical wireless systems that include Micro, Pico and Atto cells (see Figure 2a,b,c). On the ceiling of the room there are 9 cell units; and 8 of these units consist of Pico and Atto cell transmitters, while the last one includes the Micro cell transmitter. Each Pico and Atto cell unit consists of five access points (APs) and 4 of these APs provide Atto cells, while the last one provides the Pico cell. The concept of angle diversity transmitter (ADT) that has been used in this work to provide Pico and Atto cell coverage is similar to [11]. A ray tracing algorithm has been used to model the optical channel similar to [41], [42] in the simulation. The line of sight (LOS), first and second order reflections were considered in this work since there is no substantial impact on the received optical power from the third and higher order reflections [42]. Therefore, each surface (ceiling, walls and floor) inside the room was divided into identical small areas based on the order of reflection which act as small transmitters by reflecting the light rays in different directions in the shape of Lambertian patterns [5]. The dimensions of these small identical areas can affect the resolution of the results. When these identical areas become very small, a much higher resolution can be obtained. However, the computation time of the simulation increases significantly. Therefore, a reasonable identical area can be chosen to maintain a balance between the resolution of the results and the computation time of the simulation. As shown in Figure 1, the communication floor is set 1m above the floor and that means all communication links operate above the communication floor. An angle diversity receiver (ADR) that consists of seven faces similar to [11] was used, as shown in Figure 2d. Each face has a narrow Field of View (FOV) that helps collect the signal from different directions, reducing interference. The orientation of each face helps to cover a different area by utilising Azimuth ($Az$) and Elevation ($El$) angles.

Since ADT is utilised in this work and since it cannot achieve the ISO and European illumination requirements [43] based on the design used, thus, four supporting light units have been used to provide lighting only. As a result, 306.4 lx has been achieved, as shown in Figure 3, which is greater than the minimum ISO requirement and greater than the minimum European illumination requirements (300 lx). Table 1 shows the simulation parameters of the room, cells systems, transmitters and receiver.

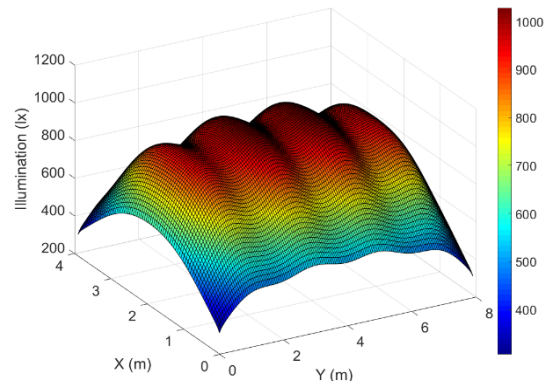

Figure 3. Illumination level inside the room.

**Table 1.** System Parameters

| Parameters | Configurations | | | | | | |
|---|---|---|---|---|---|---|---|
| **Room** | | | | | | | |
| Width × Length × Height (x, y, z) | 4 m × 8 m × 3 m | | | | | | |
| Walls and ceiling reflection coefficient | 0.8 [42] | | | | | | |
| Floor reflection coefficient | 0.3 [42] | | | | | | |
| Number of reflections | 1 | 2 | | | | | |
| Area of reflection element | 5 cm × 5 cm | 20 cm × 20 cm | | | | | |
| Order of Lambertian pattern, walls, floor and ceiling | 1 [42] | | | | | | |
| **Micro Cell Transmitter** | | | | | | | |
| Semi-angle at half power | 65º | | | | | | |
| Number of Transmitters | 1 | | | | | | |
| Transmitted optical power | 0.84 W [44] | | | | | | |
| Transmitter locations (x, y, z) | (2 m, 4 m, 3 m) | | | | | | |
| **Pico and Atto Cells Transmitters** | | | | | | | |
| Semi-angle at half power of Pico Cell | 40º | | | | | | |
| Semi-angle at half power of Atto Cell | 21º | | | | | | |
| Number of RYGB LDs per Cell | 3 | | | | | | |
| Transmitted optical power of Red, Yellow, Green and Blue Wavelengths Respectively | 0.8, 0.5, 0.3 and 0.3 W | | | | | | |
| Total transmitted power of each RYGB LD | 1.9 W | | | | | | |
| Transmitter Faces | Pico | Atto (1) | Atto (2) | Atto (3) | Atto (4) | | |
| Azimuth angles | 0º | 45º | 135º | 225º | 315º | | |
| Elevation angles | -90º | -70º | -70º | -70º | -70º | | |
| Transmitters locations (x, y, z) | (1 m, 1 m, 3 m), (1 m, 3 m, 3 m), (1 m, 5 m, 3 m), (1 m, 7 m, 3 m), (3 m, 1 m, 3 m), (3 m, 3 m, 3 m), (3 m, 5 m, 3 m) and (3 m, 7 m, 3 m) | | | | | | |
| **Supporting Light Units** | | | | | | | |
| Semi-angle at half power | 70º | | | | | | |
| Number of RYGB LDs per Light Unit | 10 | | | | | | |
| Total transmitted power of each RYGB LD | 1.9 W | | | | | | |
| Transmitters locations (x, y, z) | (2 m, 1 m, 3 m), (2 m, 3 m, 3 m), (2 m, 5 m, 3 m) and (2 m, 7 m, 3 m) | | | | | | |
| **Receiver** | | | | | | | |
| Responsivity of Infrared, Red, Yellow, Green and Blue Wavelengths Respectively | 0.6, 0.4, 0.35, 0.3 and 0.2 A/W | | | | | | |
| Number of photodetectors | 7 | | | | | | |
| Photodetector | 1 | 2 | 3 | 4 | 5 | 6 | 7 |
| Azimuth angles | 0º | 0º | 60º | 120º | 180º | 240º | 300º |
| Elevation angles | 90º | 40º | 40º | 40º | 40º | 40º | 40º |
| Field of view (FOV) of each detector | 30º | 25º | 25º | 25º | 25º | 25º | 25º |
| Area of each photodetector | 20 mm$^2$ | | | | | | |
| Receiver noise current spectral density | 4.47 pA/√Hz [5] | | | | | | |
| Receiver bandwidth | 5 GHz | | | | | | |

## 3. SIMULATION SETUP AND RESULTS

In this work, two optical transmitters are used to provide downlink communication for Micro, Pico and Atto cells. The transmitters emit infrared and visible light. The transmitter of the Micro cell produces infrared while the transmitters of the Pico and Atto cells produce visible light. laser diodes (LDs) that provide four wavelengths (Red, Yellow, Green and Blue) are used in the Pico and Atto cells as visible light sources to provide communication and illumination at the same time. These four wavelengths are aggregated to provide white light for the purpose of indoor illumination [40]. Nine cell units have been considered in this work. Each cell unit from number 1 to number 8 consists of one access point (AP) acting as a transmitter of the Pico cell and four APs acting as transmitters of the Atto cell, while cell unit number 9 consists of one AP that provides the Micro cell over the room. Wavelength division multiple access (WDMA) is used in this work to offer multiple access. Two different scenarios have been considered; each scenario consisting of 10 users. The first scenario is the worst scenario, where five users are clustered under one cell unit. The second scenario is the best case scenario where the ten users are distributed over the room. The optimisation of the resource allocation is based on maximising the sum of SINRs of all users and utilised a MILP model [13], [35]. A controller placed on the room ceiling knows the locations of the users in advance [13], [35]. The optimised resource allocation and locations of the users for both scenarios are stated in Tables 2. Figure 4 illustrates an example of three users and three APs with two wavelengths (Red and Blue) to show how the assignment worked using the MILP model. Three lines are used to differentiate the signal, noise and interference. Solid lines indicate the signal, dashed lines specify interference and dotted lines refer to the noise. In this example, the blue wavelength has been assigned to user 1 only from AP1 which shows user 1 is affected only by the noise, while the red wavelength has been assigned to users 2 and 3 from different APs (AP2 is assigned to user 2 and AP3 is assigned to user 3) which shows that the APs interfere with each other and are also affected by noise.

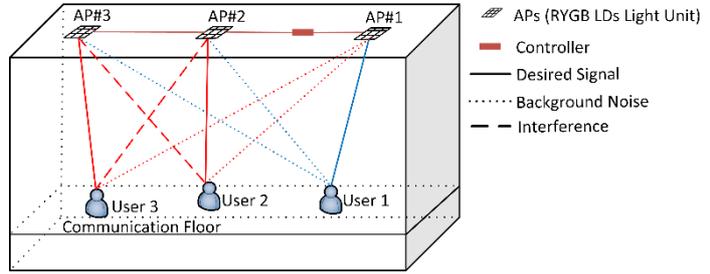

Figure 4. Example of WDMA.

**Table 2.** Scenario 1 with the optimised resource allocation

| User | Scenario 1 | | | | | | Scenario 2 | | | | | |
|---|---|---|---|---|---|---|---|---|---|---|---|---|
| | Location (x, y, z) | Cells Unit # | Cell Type | AP | Branch | Wavelength | Location (x, y, z) | Cells Unit # | Cell Type | AP | Branch | Wavelength |
| 1 | (0.5,0.5,1) | 1 | Atto | 3 | 1 | Yellow | (0.5,0.5,1) | 1 | Atto | 3 | 1 | Red |
| 2 | (0.5,1.5,1) | 1 | Atto | 2 | 1 | Yellow | (0.5,5.5,1) | 3 | Atto | 2 | 1 | Red |
| 3 | (1.0,1.0,1) | 1 | Pico | 1 | 1 | Red | (1.5,1.5,1) | 1 | Atto | 1 | 1 | Red |
| 4 | (1.5,0.5,1) | 1 | Atto | 4 | 1 | Yellow | (1.5,3.5,1) | 2 | Atto | 1 | 1 | Red |
| 5 | (1.5,1.5,1) | 1 | Atto | 1 | 1 | Yellow | (1.5,7.5,1) | 4 | Atto | 1 | 1 | Red |
| 6 | (2.5,6.5,1) | 8 | Atto | 3 | 1 | Yellow | (2.5,0.5,1) | 5 | Atto | 3 | 1 | Red |
| 7 | (2.5,7.5,1) | 8 | Atto | 2 | 1 | Yellow | (2.5,4.5,1) | 7 | Atto | 3 | 1 | Red |
| 8 | (3.0,7.0,1) | 8 | Pico | 1 | 1 | Red | (2.5,5.5,1) | 7 | Atto | 2 | 1 | Red |
| 9 | (3.5,6.5,1) | 8 | Atto | 4 | 1 | Yellow | (3.5,2.5,1) | 6 | Atto | 4 | 1 | Red |
| 10 | (3.5,7.5,1) | 8 | Atto | 1 | 1 | Yellow | (3.5,6.5,1) | 8 | Atto | 4 | 1 | Red |

After obtaining the optimized resource allocation (Cells types, APs and Wavelengths) of each user in both scenarios, the channel bandwidth and the SINR were determined at a fixed data rate.

Figure 5 shows the achievable optical channel bandwidth for each user in the two different scenarios. The optical channel bandwidth in both scenarios varies between around 7.8 GHz and 12 GHz. The user location impacts the achieved optical channel bandwidth for each user. Users close to the room corner, such as users 1 and 2 in scenario

1, and user 1 in scenario 2, have the highest optical channel bandwidth as the reflections arriving at these users are received with lower delay spread.

In Figure 6, the SINR has been evaluated at a fixed data rate of 7.1 Gbps for each user in the two scenarios. In Scenario 2, where ten users are distributed over the room, all users are assigned to Atto cells and the SINR is higher compared to Scenario 1, where five users are clustered under one cell unit. In scenario one, users who are assigned to Atto cells (users 1, 2, 4, 5, 6, 7, 9 and 10) provide a better SINR compared to users assigned to the Pico cell (users 3 and 8). However, all users in the two scenarios offer a higher level of SINR than the threshold (15.6 dB) required for bit error rate of $10^{-9}$ when using On Off Keying (OOK) modulation.

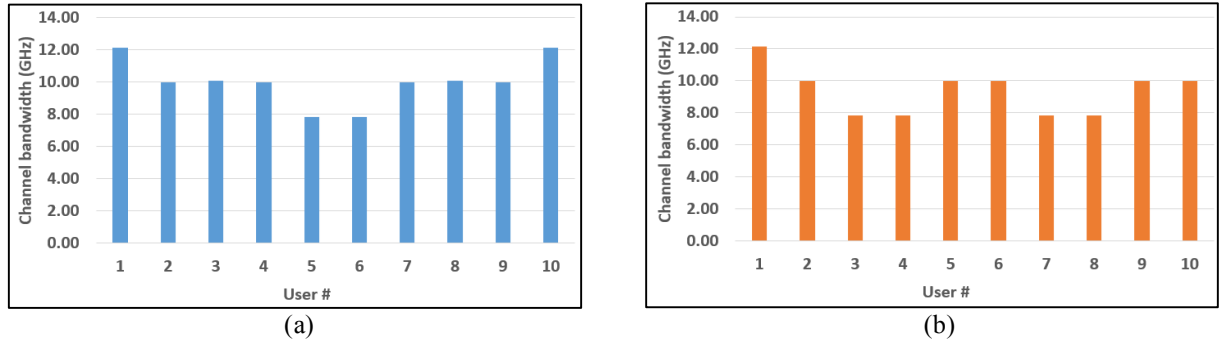

Figure 5. Channel bandwidth (a) Scenario #1, (b) Scenario #2.

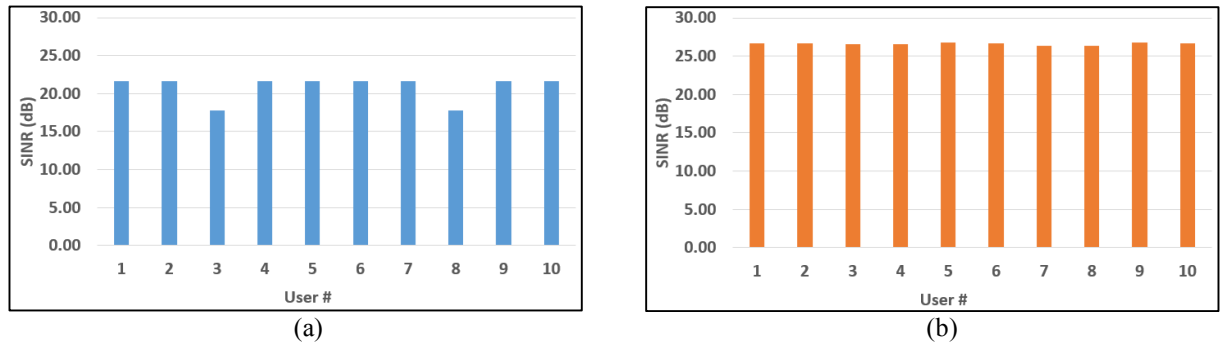

Figure 6. SINR (a) Scenario #1, (b) Scenario #2.

## 4. CONCLUSIONS

This paper investigated the impact of resource allocation in co-existing optical wireless HetNets. Two different scenarios, each consisting of ten users, have been considered in this work. A mixed-integer linear programming (MILP) model is developed to establish the optimum resource allocation in a proposed wavelength division multiple access (WDMA) optical wireless system. The optimum allocation of wavelengths and access points (APs) to each user to support multiple users in an environment where Micro, Pico and Atto Cells co-exist for downlink communication has been considered in this work. Visible light links which utilise four wavelengths per access point (red, green, yellow and blue) have been used for Pico and Atto cell systems, while the Micro cell system used an infrared (IR) transmitter. An ADR consisting of seven branches was evaluated in this work. All users in both scenarios achieved a high optical channel bandwidth beyond 7.8 GHz. In addition, all users in the two scenarios achieved a high SINR beyond the threshold (15.6 dB) at a data rate of 7.1 Gbps.

**ACKNOWLEDGEMENTS**

The authors would like to acknowledge funding from the Engineering and Physical Sciences Research Council (EPSRC) INTERNET (EP/H040536/1), STAR (EP/K016873/1) and TOWS (EP/S016570/1) projects. The authors extend their appreciation to the deanship of Scientific Research under the International Scientific Partnership Program ISPP at King Saud University, Kingdom of Saudi Arabia for funding this research work through ISPP#0093. OZA would like to thank Umm Al-Qura University in the Kingdom of Saudi Arabia for funding his PhD scholarship, SOMS would like to thank the University of Leeds and the Higher Education Ministry in Sudan for funding her PhD scholarship. SHM would like to thank EPSRC for providing her Doctoral Training Award scholarship. All data are provided in full in the results section of this paper.